\documentclass[journal=jpclcd,manuscript=letter]{achemso}
\setkeys{acs}{articletitle=true}

\usepackage{amssymb}
\usepackage{amsmath}
\usepackage{xcolor}
\usepackage{braket}
\usepackage{cancel}
\usepackage{amsfonts}
\usepackage{graphicx}
\usepackage{mathptmx}
\usepackage{etoolbox}
\usepackage{multirow}
\usepackage{makecell}

\author{Jakub Lang}
\author{Micha\l\ Przybytek}
\author{Micha\l\ Lesiuk}
\email{m.lesiuk@uw.edu.pl}
\affiliation{University of Warsaw, Faculty of Chemistry, Warsaw, Poland}
\date{\today}

\title{Estimating complete basis set extrapolation error through random walk}

\begin{document}

\begin{tocentry}
\includegraphics[width=\linewidth]{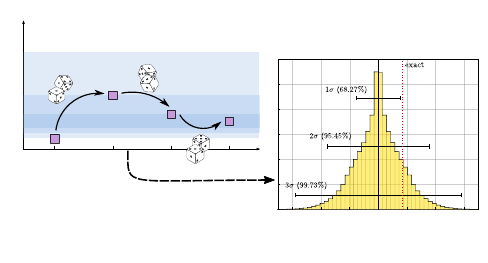}
\end{tocentry}

\begin{abstract}
We propose a method of estimating the uncertainty of a result obtained through extrapolation to the complete basis set limit. The method is based on an ensemble of random walks which simulate all possible extrapolation outcomes that could have been obtained if results from larger basis sets had been available. The results assembled from a large collection of random walks can be then analyzed statistically, providing a route for uncertainty prediction at a confidence level required in a particular application. The method is free of empirical parameters and compatible with any extrapolation scheme. The proposed technique is tested in a series of numerical trials by comparing the determined confidence intervals with reliable reference data. We demonstrate that the predicted error bounds are reliable, tight, yet conservative at the same time.
\end{abstract}

The demand for accurate quantum-chemical calculations for many-electron atoms and molecules has been rapidly increasing in recent years, fueled by developments in the fields such as ultracold chemistry and physics~\cite{tomza19,gronowski20,smialkowski21,ladjimi23,finelli24}, quantum-based metrology~\cite{lesiuk20b,hellmann22,lang23,lang2023collision,lesiuk23,garberoglio23},
spectroscopy~\cite{lesiuk19,lesiuk20,gebala23,landau23,isolde25} or search for effects beyond the standard model~\cite{safronova18,gaul19,tomza21,marc23,chamorro24}. It is striking that in a vast majority of these studies, it is not only necessary to provide accurate theoretical results that account for all relevant physical effects, but also estimate the uncertainty of the calculated data. Simultaneously, most calculations of this type employ a basis set for expansion of spinorbitals/spinors which naturally leads to an error that must be controlled. It is well-known that due to the electronic cusp condition~\cite{kato57,helgaker00}, results of correlated calculations converge slowly with respect to the basis set size. Consequently, development of methods that reduce the basis set incompleteness error remains an active field of research. Explicitly correlated methods~\cite{haettig12,kong12,tenno12}, transcorrelated approaches~\cite{boiii69}, density-based corrections~\cite{loos19a,loos19b,giner20,mester25}, and extrapolation techniques~\cite{klopper01,varandas07,feller11,varandas18} are frequently applied for this purpose. In this paper we focus on the last family of methods.

Extrapolation to the complete basis set limit is an attractive option of reducing the basis set incompleteness error due to its conceptual simplicity, vanishingly small computational cost and broad applicability. 
Several extrapolation methods are frequently used in the literature and there is general consensus that, when used with care, they considerably improve the results (see, for example, Ref.~\citenum{feller11} for a detailed analysis). 
However, estimation of uncertainty of the extrapolated results and determination of proper error bars are challenging issues with no general guidelines available. 
Assignment of the uncertainty is usually based on, for example, comparing extrapolated results from a progression of basis sets~\cite{gebala23,lesiuk23}, applying different extrapolation schemes and observing variation between them~\cite{guo24,chamorro24}, or comparing the extrapolated result with the value obtained with the largest available basis set~\cite{gronowski20,lang24}.
Alternatively, comparison with external reference data, either theoretical or experimental, is an option for selecting the proper extrapolation protocol, but such data may not be available in many situations. 
In any case, estimation of the residual extrapolation error frequently involves a degree of arbitrariness or secondary assumptions.

Another problem related to this issue, which is particularly important at the interface of theory and experiment, is a different meaning of the uncertainties in these fields. 
In the experiment, one typically repeats the same measurement numerous times and assumes that the variation in the data is represented by a certain probability distribution. 
The uncertainties are then assigned based on confidence intervals resulting from this distribution, leading to a clear statistical meaning of the error bars. 
Such procedure is usually impossible in theory and hence the meaning of the error bars assigned to a theoretical result is simply a statement that with a sufficiently high probability the exact result differs from the calculation by less than a certain value. 
However, it is typically not known what this probability really is and there is no way of tracing it back to any confidence interval based on statistical analysis. 
Of course, it is also possible to compare various extrapolation schemes by benchmarking against a set of reference data~\cite{feller11}, but there is no guarantee that the conclusions can be transferred to a particular problem at hand which is outside the training set. In other words, this approach is inherently not system-specific.

In this work we propose a method of assigning uncertainties to theoretical results obtained by extrapolation. The method is based on a series of random walks which simulate possible results that could have been obtained if data calculated with larger basis sets had been available. While a single random walk does not carry any practical information, an ensemble of random walks can be analyzed statistically to uncover the possible variation in the extrapolated results. This provides a route for uncertainty prediction without arbitrary assumptions in a system-specific way.

In order to introduce the proposed method, let us consider calculation of a certain quantity $E$ using a progression of basis sets\bibnote{While the symbol $E$ is used to denote the quantity of interest throughout the paper, it does not imply that we are referring only to energy.} and subsequent extrapolation of the results to the complete basis set limit. The size of the basis set is denoted by a single parameter $X$ (for example, the cardinality in the case of correlation-consistent basis sets~\cite{dunning89}). The value of $E$ calculated within the basis set $X$ is denoted by the symbol $E_X$. For sufficiently large $X$ we can expand $E_X$ in the asymptotic series:
\begin{align}
\label{exasym}
    E_X = E_\infty + \sum_{n=3}^\infty \frac{A_n}{X^n}.
\end{align}
It is well-known that in the case of electronic energy and many other quantities, the dominant term of this expansion is proportional to $X^{-3}$ (see Refs.~\citenum{carroll79,hill85,kutzelnigg92,kutzelnigg08}). Many extrapolation procedures use this information, either directly or implicitly. In this work we shall employ primarily the two-point extrapolation scheme of Helgaker~\emph{et al.}~\cite{helgaker97,halkier98} which is based on truncating the above expression after the leading-order term, i.e., $E_X = E_\infty + \frac{A_3}{X^3}$. Next, the results obtained with two consecutive basis sets, $E_X$ and $E_{X-1}$, are combined to eliminate the $A_3$ coefficient. This gives the following explicit formula for the estimate of the complete basis set limit:
\begin{align}
\label{helgaker}
    E_\infty \approx \frac{E_X\,X^3 - E_{X-1}(X-1)^3}{X^3 - (X-1)^3}.
\end{align}
Let us denote the value extrapolated according to Eq.~(\ref{helgaker}) from the pair of basis sets $(X,X-1)$ by the symbol $e_X$.

Estimation of the extrapolation error is a difficult task primarily because (i) little is known analytically about the higher-order terms in Eq.~(\ref{exasym}) and the coefficients $A_n$ for many-electron systems, (ii) the accessible range of $X$ is typically too narrow to determine them reliably, e.g., by fitting, (iii) secondary sources of error such as radial incompleteness may play a role for any finite $X$. In this work we adopt a minimalist assumption about the behavior of $e_X$ as a function of $X$. We assume only that the absolute differences between neighboring extrapolated values, $|e_X - e_{X-1}|$, decrease monotonically for sufficiently large $X$, but $e_X$ themselves do not need to follow any consistent pattern. For example, in the case of the extrapolation formula~(\ref{helgaker}), one can show that these differences behave for large $X$ as
\begin{align}
\label{eqbound}
    e_X - e_{X-1} = \frac{\mathcal{C}}{X^5} + \ldots,
\end{align}
where $\mathcal{C}$ is a system-dependent numerical constant and the higher-order terms (proportional to $X^{-n}$ with $n\geq 6$) are not written explicitly. From this formula it is evident that the quantities $|e_X - e_{X-1}|$ decrease monotonically for sufficiently large $X$, even if $e_X$ themselves do not exhibit a monotonic behavior, e.g., oscillate. Note that the value of $\mathcal{C}$ could, in principle, be obtained by using results from a progression of basis sets, but we found that such approach is not trustworthy when applied within the range of $X$ that is typically available.

Let us assume we carried out calculations within three consecutive basis sets: $X$, $X-1$, and $X-2$, while results for larger basis sets are not available. From this data we can assemble two extrapolated values, $e_X$ and $e_{X-1}$. According to our main assumption, if the next extrapolated value ($e_{X+1}$) had been available, it would have been bounded by:
\begin{align}
\label{bound1}
    e_X - |e_X - e_{X-1}| < e_{X+1} < e_X + |e_X - e_{X-1}|.
\end{align}
At face value, this inequality in itself is not very useful, because we do not know what the actual value of $e_{X+1}$ is. More importantly, there is no guarantee that the exact result ($E_\infty$) also lies within this interval. However, we can pessimistically assume that any value of $e_{X+1}$ within the bounding interval is equally probable and randomize it from a uniform distribution. In this way we obtain a value of $\tilde{e}_{X+1}$ which represents one possible scenario of what the actual $e_{X+1}$ may be. This procedure is then continued. Assuming the randomized value of $\tilde{e}_{X+1}$ we know that the next extrapolated value ($e_{X+2}$) is bounded by:
\begin{align}
\label{bound2}
    \tilde{e}_{X+1} - |\tilde{e}_{X+1} - e_X| < e_{X+2} < \tilde{e}_{X+1} + |\tilde{e}_{X+1} - e_X|,
\end{align}
and again randomize $\tilde{e}_{X+2}$ within this interval. This procedure eventually converges in the sense that after a certain number of steps $N$, the length of the bounding interval becomes smaller than a predefined threshold. At the same time, two successive randomized values ($\tilde{e}_{X+N}$ and $\tilde{e}_{X+N-1}$) obviously differ by less than this threshold. In the following, we refer to the set of $\tilde{e}_{X+1}$, $\tilde{e}_{X+2}$, $\ldots$ as a trajectory and denote the converged value $\tilde{e}_{X+N}$ by $\tilde{e}_\infty$.

A single trajectory in the proposed method is essentially a random walk, where the values of $\tilde{e}_{X+1}$, $\tilde{e}_{X+2}$, $\ldots$ are allowed to randomly shift within the corresponding bounding intervals. However, we stress that a single trajectory obtained in this way is not useful for any practical purpose. It represents only one possible scenario of what could have happened if results in larger basis sets had been available (allowing to obtain the subsequent extrapolated results $e_{X+1}$, $e_{X+2}$, $\ldots$). The proposed method becomes useful only when a large number of trajectories is run independently. It provides an insight into the variability of $\tilde{e}_\infty$ without any assumptions about the particular values of $e_{X+1}$, $e_{X+2}$, $\ldots$. The only assumption used in this procedure is the monotonic decrease of the absolute differences between extrapolated values as a function of $X$. Having a large number of $\tilde{e}_\infty$ obtained from separate trajectories, the results can be analyzed statistically. This naturally leads to system-specific uncertainty estimates for the average value of $\tilde{e}_\infty$, as demonstrated further in the text.

Let us first illustrate the proposed method by applying it to two model systems for which both reliable reference data and results obtained within a progression of basis sets are available. Our main goal here is a detailed discussion of the algorithm of the proposed method, while presentation of results for a much larger set of systems is given later. The first example is the electronic correlation energy of the H$_2$ molecule (internuclear distance $1.4\,$a.u.) calculated within aug-mcc-pVXZ basis sets of Mielke~\emph{et al.}~\cite{mielke99} using the full configuration interaction (FCI) method. The reference values for the total and Hartree-Fock energies of H$_2$ come from papers of Pachucki~\cite{pachucki10} and Mitin~\cite{mitin00}, respectively, giving the near-exact value of the correlation energy equal to $-40.846\,348\,$mHa. The second example is the correlation energy of the carbon atom calculated at the FCI/aug-cc-pCVXZ~\cite{kendall92} level of theory. Based on accurate results for the total energy obtained by Strasburger~\cite{strasburger19} and the Hartree-Fock energy by Bunge~\emph{et al.}~\cite{bunge93}, the reference value for the correlation energy is $-156.287\,$mHa. In the first example, results within basis sets up to $X=6$ are available, while for the second example we are limited to $X=4$. The test cases were purposefully chosen to study the performance of the proposed method in these two distinct situations, both of which are frequently encountered in practice. The raw results used in our analysis were calculated in Ref.~\citenum{lesiuk19b} and are reproduced in Table~\ref{tab:tests12} for convenience.

\begin{table}
\caption{\label{tab:tests12}
Raw data and summary of the results for two selected test cases (see text). All values are given in mHa (with signs reversed for convenience).
}
\begin{tabular}{ccccc}
\hline\hline
 & \multicolumn{2}{c}{Test case 1} & \multicolumn{2}{c}{Test case 2} \\
\hline
$X$ & $-E_X$ & $-e_X$ & $-E_X$ & $-e_X$ \\
\hline
2 & ---     & ---     & 132.539 & --- \\
3 & ---     & ---     & 145.934 & 151.574 \\
4 & 40.6528 & ---     & 151.029 & 154.747 \\
5 & 40.7374 & 40.8262 & ---     & --- \\
6 & 40.7797 & 40.8378 & ---     & --- \\
\hline
\multicolumn{5}{c}{Best estimates} \\
\hline
$1\sigma\,$(68.27\%) & \multicolumn{2}{c}{40.8378 $\pm$ 0.0078} & \multicolumn{2}{c}{154.7 $\pm$ 2.1} \\
$2\sigma\,$(95.45\%) & \multicolumn{2}{c}{40.838 $\pm$ 0.018} & \multicolumn{2}{c}{154.7 $\pm$ 4.8} \\
$3\sigma\,$(99.73\%) & \multicolumn{2}{c}{40.838 $\pm$ 0.029} & \multicolumn{2}{c}{154.7 $\pm$ 7.9} \\
\hline
true error$^a$ & 
\multicolumn{2}{c}{0.0085} & \multicolumn{2}{c}{1.540} \\
\hline
reference & \multicolumn{2}{c}{40.8463} & \multicolumn{2}{c}{156.287} \\
\hline\hline
\multicolumn{5}{l}{$^a$absolute deviation from the reference data given in the last row}
\end{tabular}
\end{table}

The random walk procedure was initiated using the extrapolated value from the pair of two largest basis sets. However, results from three consecutive basis sets are necessary to establish the initial bounding interval, see Eq.~(\ref{bound1}). About ten million trajectories were simulated in both test examples; further increase of this parameter leads to no appreciable changes in the uncertainty predictions. The values of $\tilde{e}_{X}$ were randomized from a uniform distribution. Each random walk was stopped when the width of the bounding interval, see Eqs.~(\ref{bound1})~and~(\ref{bound2}), falls below the threshold of $10^{-16}$. The converged values $\tilde{e}_\infty$ for each trajectory were recorded and are the subject of the analysis that follows.

\begin{figure}
    \centering\hspace{-1.0cm}
    \includegraphics[scale=0.60]{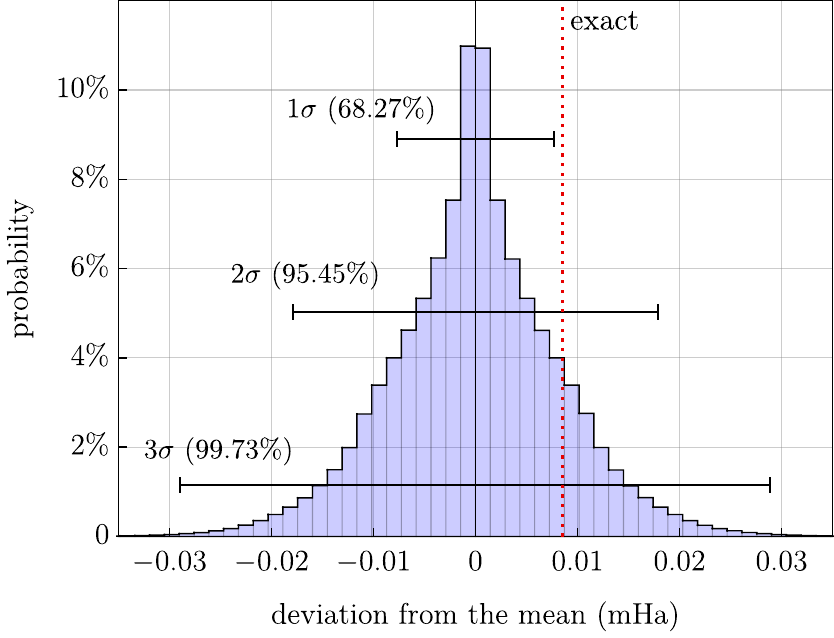}\\
    \vspace{0.5cm}\hspace{-1.0cm}
    \includegraphics[scale=0.60]{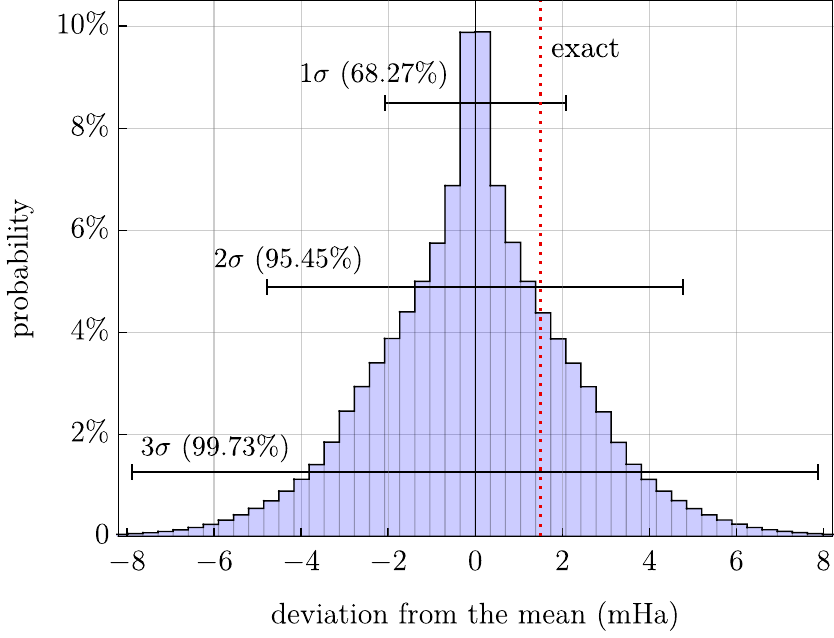}
    \caption{Histograms illustrating the results of about $10^7$ random walks for test case 1 (upper panel) and test case 2 (lower panel). The histograms are centered such that the sample mean corresponds to zero at the horizontal axis. The deviations from the mean are given in mHa. The $1\sigma$, $2\sigma$, and $3\sigma$ confidence intervals (see text) are shown as overlaying brackets. The reference (near-exact) values are represented as red dotted lines.}
    \label{fig:histograms12}
\end{figure}

In Fig.~\ref{fig:histograms12} we provide histograms illustrating the distribution of $\tilde{e}_\infty$ obtained after about $10^7$ random walks. The distributions are nearly symmetric with respect to the mean which is not surprising considering that the endpoints of the bounding intervals, see Eqs.~(\ref{bound1})~and~(\ref{bound2}), are always equidistant from the previous extrapolated value. For the same reason, as the sample size increases the average value of $\tilde{e}_\infty$ obtained from all walks should converge to the extrapolated result $e_X$ which was used to initiate the random walks, see Eq.~(\ref{bound1}) and the accompanying discussion. This is confirmed in our calculations, with agreement of six significant digits in all cases. Therefore, we reiterate that the method proposed in this work enables us to estimate the uncertainty of an extrapolated result, while the result itself is unchanged in comparison with the value $e_X$ used to initiate the random walks, see Table~\ref{tab:tests12}.

The probability distributions represented in Fig.~\ref{fig:histograms12} enable us to assign confidence intervals to the extrapolated results. We consider three confidence intervals at the confidence level of $68.27\%$, $95.45\%$, and $99.73\%$. The choice of these percentages is arbitrary and is motivated by analogy to the commonly used values in the case of the normal distribution. However, we stress that the probability distributions obtained in the present context are clearly not normal and hence the lengths of the confidence intervals are not simple multiples of the standard deviation calculated from the sample, as in the case of the Gaussian distribution. Instead, the confidence intervals are defined as intervals centered at the sample mean which cover a given percentage of the data points, as illustrated in Fig.~\ref{fig:histograms12}. For brevity and by analogy with the normal distribution, we refer to the confidence intervals at the confidence level of $68.27\%$, $95.45\%$, and $99.73\%$ by $1\sigma$, $2\sigma$, and $3\sigma$, respectively. The confidence intervals determined by the proposed procedure for H$_2$ molecule (test case 1) and carbon atom (test case 2) are shown in Table~\ref{tab:tests12}. In both cases, they successfully estimate the extrapolation error. In the former case, $2\sigma$ confidence interval correctly predicts the deviation from the reference value, while in the latter even the $1\sigma$ confidence interval is sufficient for this purpose.

As a side note, we mention that according to the numerical tests, the probability distributions shown in Fig.~\ref{fig:histograms12} do not seem to be well represented by a simple analytic form such as Laplace (bivariate exponential) distribution. We were not able to find the exact analytic form of this distribution in the limit of infinite number of independent trajectories. Mathematically this is a difficult task, because the randomization steps involved in a single trajectory are strongly interdependent, i.e., the interval in which the subsequent randomization is performed depends directly on the result of previous two samplings. From a pragmatic standpoint, the lack of this information is not problematic, because the computational cost of running a single trajectory is very low. Therefore, assembling a sufficient number of samples for a credible statistical analysis is not challenging: calculations with ten million random walks take mere seconds.

To illustrate the performance of the proposed method for a larger set of examples, we gathered numerous results from the literature where results of the calculations from a progression of basis sets is available and, simultaneously, reliable reference data is found. The main source of the reference values are either explicitly correlated calculations (explicitly-correlated Gaussians~\cite{szalewicz10,mitroy13} or F12 methodology~\cite{haettig12,kong12,tenno12}) or calculations with significantly larger basis sets than used in the error estimation procedure. The benchmark set includes both correlation energies, given in Table~\ref{tab:tests_corr}, and other quantities such as atomization energies, interaction energies, or polarizabilities, given in Table~\ref{tab:tests_prop}. In all cases, around ten million trajectories were run which is sufficient to make the confidence intervals stable to all digits shown (as a rule, the last digit has always been rounded up).

\begin{table*}
\caption{\label{tab:tests_corr}
Estimated extrapolation errors for correlation energies of a benchmark set of systems. A brief description of the data and level of theory are given in the first and second columns, respectively. The maximum cardinal number, $X_\mathrm{max}$, used in the procedure is given in the third column. The determined error bars at the $1\sigma$, $2\sigma$, and $3\sigma$ confidence levels (see text for precise definitions) are given in the fifth, sixth and seventh columns, respectively. The reference result is given in the last column, while deviation of a given result from the corresponding reference data in the second last column. The most narrow confidence interval which correctly predicts the difference from the reference result is shown in bold. All results are given in mHa.}
{\footnotesize
\addtolength{\tabcolsep}{-0.2em}
\begin{tabular}{cccccccccc}
\hline\hline
  & method  & & & \multicolumn{3}{c}{confidence intervals} &  \\
 quantity & and basis & $X_\mathrm{max}$ & $e_{X_\mathrm{max}}$ & 
$1\sigma$ & $2\sigma$& $3\sigma$ & 
 error$^a$ &
 reference value \\
 & & & & (68.27\%) & (95.45\%) & (99.73\%) & & \\
\hline
\multirow{4}{*}{\makecell{He atom \\ correlation energy}} &
\multirow{4}{*}{\makecell{FCI \\ d$X$Z, Ref.~\citenum{cencek12} }} 
    & 4 & $-$41.907 & $\pm${\bf 0.17} & $\pm$0.37 & $\pm$0.62 & 0.137 &
\multirow{4}{*}{\makecell{$-$42.044 381 \\ Refs.~\citenum{lehtola19,nakashima08}}} \\
 &  & 5 & $-$41.983 & $\pm$0.050 & $\pm${\bf 0.12} & $\pm$0.19 & 0.062 \\
 &  & 6 & $-$42.013 & $\pm$0.020 & $\pm${\bf 0.045} & $\pm$0.074 & 0.032 \\
 &  & 7 & $-$42.026 & $\pm$0.009 & $\pm${\bf 0.021} & $\pm$0.034 & 0.018 \\
\hline
\multirow{3}{*}{\makecell{Be atom \\ correlation energy}} &
\multirow{3}{*}{\makecell{FCI \\ Slater-type basis \\ Ref.~\citenum{lesiuk19} }} 
    & 4 & $-$94.099 & $\pm$0.20 & $\pm${\bf 0.46} & $\pm$0.76 & 0.233 &
\multirow{3}{*}{\makecell{$-$94.332 459 \\ Ref.~\citenum{pachucki04}}} \\
 &  & 5 & $-$94.253 & $\pm${\bf 0.11} & $\pm$0.23 & $\pm$0.39 & 0.079 \\
 &  & 6 & $-$94.305 & $\pm${\bf 0.035} & $\pm$0.078 & $\pm$0.13 & 0.027 \\
\hline
\multirow{2}{*}{\makecell{Be atom \\ correlation energy}} &
\multirow{2}{*}{\makecell{MP2 \\ aug-cc-pwCV$X$Z, Ref.~\citenum{prascher11} }} 
    & 4 & $-$75.711 & $\pm${\bf 1.2} & $\pm$2.7 & $\pm$4.4 & 0.648 &
\multirow{2}{*}{\makecell{$-$76.358 \\ Ref.~\citenum{przybytek18}}} \\
 &  & 5 & $-$76.085 & $\pm$0.25 & $\pm${\bf 0.56} & $\pm$0.93 & 0.274 \\
\hline
\multirow{2}{*}{\makecell{Be atom \\ correlation energy}} &
\multirow{2}{*}{\makecell{CCSD \\ aug-cc-pwCV$X$Z, Ref.~\citenum{prascher11} }} 
    & 4 & $-$93.633 & $\pm${\bf 0.59} & $\pm$1.4 & $\pm$2.2 & 0.031 &
\multirow{2}{*}{\makecell{$-$93.665 \\ Ref.~\citenum{przybytek18}}} \\
 &  & 5 & $-$93.586 & $\pm$0.032 & $\pm$0.071 & $\pm${\bf 0.12} & 0.079 \\
\hline
\multirow{2}{*}{\makecell{H$_3^+$ cation \\ correlation energy}} &
\multirow{2}{*}{\makecell{FCI \\ aug-mcc-pV$X$Z, Ref.~\citenum{mielke99} }} 
    & 4 & $-$43.432 & $\pm${\bf 0.062} & $\pm$0.14 & $\pm$0.24 & 0.032 &
\multirow{2}{*}{\makecell{$-$43.464 \\ Refs.~\citenum{pavanello09} and~\citenum{jensen05}}} \\
 &  & 5 & $-$43.441 & $\pm$0.007 & $\pm$0.014 & $\pm${\bf 0.024} & 0.023 \\
\hline
\multirow{2}{*}{\makecell{LiH molecule \\ correlation energy}} &
\multirow{2}{*}{\makecell{MP2 \\ aug-cc-pwCV$X$Z, Ref.~\citenum{prascher11} }} 
    & 4 & $-$72.343 & $\pm${\bf 1.5} & $\pm$3.2 & $\pm$5.3 & 0.546 &
\multirow{2}{*}{\makecell{$-$72.890 \\ Refs.~\citenum{bukowski99}}} \\
 &  & 5 & $-$72.660 & $\pm$0.21 & $\pm${\bf 0.48} & $\pm$0.79 & 0.230 \\
\hline
\multirow{2}{*}{\makecell{LiH molecule \\ correlation energy}} &
\multirow{2}{*}{\makecell{CCSD \\ aug-cc-pwCV$X$Z, Ref.~\citenum{prascher11} }} 
    & 4 & $-$83.103 & $\pm${\bf 0.27} & $\pm$0.60 & $\pm$1.0 & 0.113 &
\multirow{2}{*}{\makecell{$-$82.990 \\ Refs.~\citenum{bukowski99}}} \\
 &  & 5 & $-$82.623 & $\pm$0.32 & $\pm${\bf 0.72} & $\pm$1.2 & 0.367 \\
\hline
\multirow{6}{*}{\makecell{Ne atom \\ correlation energy}} &
\multirow{6}{*}{\makecell{frozen-core (1s$^2$) \\ MP2/$X$ZaP \\ Refs.~\citenum{barnes08,barnes10}}} 
    & 4 & $-$315.628 & $\pm${\bf13 } & $\pm$28 & $\pm$46 & 4.595 &
\multirow{6}{*}{\makecell{$-$320.223 \\ Refs.~\citenum{flores93,flores08}}} \\
 &  & 5 & $-$319.003 & $\pm${\bf2.3 } & $\pm$5.1 & $\pm$8.4 & 1.220 & \\
 &  & 6 & $-$319.600 & $\pm$0.40 & $\pm${\bf 0.90} & $\pm$1.5 & 0.622 & \\
 &  & 7 & $-$319.881 & $\pm$0.19 & $\pm${\bf 0.42} & $\pm$0.70 & 0.342 & \\
 &  & 8 & $-$319.985 & $\pm$0.070 & $\pm$0.16 & $\pm${\bf 0.26} & 0.238 & \\
 &  & 9 & $-$320.073 & $\pm$0.059 & $\pm$0.14 & $\pm${\bf 0.22} & 0.150 & \\
\hline 
\multirow{5}{*}{\makecell{Ne atom \\ correlation energy}} &
\multirow{5}{*}{\makecell{(T) correction \\ $X$ZaP basis \\ Refs.~\citenum{barnes08,barnes10}}} 
    & 4 & $-$6.535 & $\pm${\bf 0.60} & $\pm$1.4 & $\pm$2.3 & 0.038 &
\multirow{5}{*}{\makecell{$-$6.497 \\ Ref.~\citenum{barnes08}}} \\
 &  & 5 & $-$6.647 & $\pm$0.075 & $\pm${\bf 0.17} & $\pm$0.28 & 0.150 & \\
 &  & 6 & $-$6.554 & $\pm${\bf 0.062} & $\pm$0.14 & $\pm$0.24 & 0.057 & \\
 &  & 7 & $-$6.530 & $\pm$0.016 & $\pm${\bf 0.035} & $\pm$0.058 & 0.033 & \\
 &  & 8 & $-$6.518 & $\pm$0.008 & $\pm$0.018 & $\pm${\bf 0.030} & 0.021 & \\
\hline
\multirow{3}{*}{\makecell{H$_2$O molecule \\ correlation energy}} &
\multirow{3}{*}{\makecell{MP2 \\ cc-pV$X$Z \\ Ref.~\citenum{dunning89} }} 
    & 4 & $-$298.39 & $\pm${\bf 7.8} & $\pm$18 & $\pm$29 & 1.96 &
\multirow{3}{*}{\makecell{$-$300.35 \\ Ref.~\citenum{klopper01}}} \\
 &  & 5 & $-$300.67 & $\pm${\bf 1.6} & $\pm$3.4 & $\pm$5.7 & 0.32 \\
 &  & 6 & $-$300.29 & $\pm${\bf 0.26} & $\pm$0.57 & $\pm$0.95 & 0.06 \\
\hline
\multirow{3}{*}{\makecell{CH$_2$ molecule \\ correlation energy}} &
\multirow{3}{*}{\makecell{MP2 \\ cc-pV$X$Z \\ Ref.~\citenum{dunning89} }} 
    & 4 & $-$155.08 & $\pm${\bf 3.0} & $\pm$6.7 & $\pm$11 & 0.73 &
\multirow{3}{*}{\makecell{$-$155.81 \\ Ref.~\citenum{klopper01}}} \\
 &  & 5 & $-$155.62 & $\pm${\bf 0.36} & $\pm$0.81 & $\pm$1.4 & 0.19 \\
 &  & 6 & $-$155.73 & $\pm${\bf 0.08} & $\pm$0.17 & $\pm$0.28 & 0.08 \\
\hline
\multirow{2}{*}{\makecell{HF molecule \\ correlation energy}} &
\multirow{2}{*}{\makecell{CCSD (singlet pairs) \\ cc-pV$X$Z, Ref.~\citenum{dunning89} }} 
    & 5 & $-$213.72 & $\pm${\bf 0.61} & $\pm$1.4 & $\pm$2.3 & 0.58 &
\multirow{2}{*}{\makecell{$-$213.14 \\ Ref.~\citenum{klopper01}}} \\
 &  & 6 & $-$213.34 & $\pm${\bf 0.26} & $\pm$0.57 & $\pm$0.95 & 0.20 \\
\hline
\multirow{2}{*}{\makecell{F$_2$ molecule \\ correlation energy}} &
\multirow{2}{*}{\makecell{CCSD (singlet pairs) \\ cc-pV$X$Z, Ref.~\citenum{dunning89} }} 
    & 5 & $-$414.83 & $\pm${\bf 1.9} & $\pm$4.2 & $\pm$6.9 & 0.67 &
\multirow{2}{*}{\makecell{$-$414.16 \\ Ref.~\citenum{klopper01}}} \\
 &  & 6 & $-$414.44 & $\pm$0.26 & $\pm${\bf 0.58} & $\pm$0.97 & 0.28 \\
 \hline\hline
 \multicolumn{9}{l}{$^a$absolute error with respect to the reference value given in the last column}\\
\end{tabular}
}
\end{table*}

\begin{table*}
\caption{\label{tab:tests_prop}
Same as Table~\ref{tab:tests_corr} but for properties other than atomic/molecular energies. The units are given in the first column in each case.}
{\footnotesize
\addtolength{\tabcolsep}{-0.4em}
\begin{tabular}{cccccccccc}
\hline\hline
  & method  & & & \multicolumn{3}{c}{confidence intervals} &  \\
 quantity & and basis & $X_\mathrm{max}$ & $e_{X_\mathrm{max}}$ & 
 $1\sigma$& $2\sigma$ & $3\sigma$& 
 error$^a$ &
 reference value \\
  & & & & (68.27\%) & (95.45\%) & (99.73\%) & & \\
\hline
\multirow{3}{*}{\makecell{HF molecule \\ atomization energy \\ (in kJ/mol)}} &
\multirow{3}{*}{\makecell{CCSD(T) \\ aug-cc-pCV$X$Z}} 
    & 5 & 187.54 & $\pm${\bf 0.94} & $\pm$2.1 & $\pm$3.5 & 0.23 &
\multirow{3}{*}{\makecell{187.32 $\pm$0.13 \\ Ref.~\citenum{thorpe21}}} \\
 &  & 6 & 187.35 & $\pm${\bf 0.13} & $\pm$0.30 & $\pm$0.49 & 0.03 & \\
 &  & 7 & 187.33 & $\pm${\bf 0.01} & $\pm$0.03 & $\pm$0.04 & 0.01 \\
\hline
\multirow{3}{*}{\makecell{N$_2$ molecule \\ atomization energy \\ (in kJ/mol)}} &
\multirow{3}{*}{\makecell{CCSD(T) \\ aug-cc-pCV$X$Z}} 
    & 5 & 472.05 & $\pm$0.54 & $\pm$1.2 & $\pm${\bf 2.0} & 1.54 &
\multirow{3}{*}{\makecell{470.51 $\pm$ 0.10 \\ Ref.~\citenum{thorpe21}}} \\
 &  & 6 & 471.26 & $\pm$0.53 & $\pm${\bf 1.2} & $\pm$2.0 & 0.75 & \\
 &  & 7 & 470.99 & $\pm$0.19 & $\pm$0.41 & $\pm${\bf 0.68} & 0.48 & \\
\hline
\multirow{3}{*}{\makecell{helium dimer \\ int. energy (in K) \\ internuclear dist. $5.6\,$a.u.}} &
\multirow{3}{*}{\makecell{FCI \\ d$X$Z, Refs.~\citenum{cencek12,przybytek17} }} 
    & 5 & 11.097 & $\pm${\bf 0.20} & $\pm$0.45 & $\pm$0.75 & 0.096 &
\multirow{3}{*}{\makecell{11.001 \\ Refs.~\citenum{cencek12,przybytek17}}} \\
 &  & 6 & 10.986 & $\pm${\bf 0.074} & $\pm$0.17 & $\pm$0.28 & 0.015 \\
 &  & 7 & 10.968 & $\pm$0.012 & $\pm$0.027 & $\pm${\bf 0.045} & 0.033 \\
\hline
\multirow{3}{*}{\makecell{helium dimer \\ int. energy (in K) \\ internuclear dist. $3.0\,$a.u.}} &
\multirow{3}{*}{\makecell{FCI \\ d$X$Z, Refs.~\citenum{cencek12,przybytek17} }} 
    & 5 & 3771.15 & $\pm${\bf 3.8} & $\pm$8.4 & $\pm$14 & 3.78 &
\multirow{3}{*}{\makecell{ 3767.73 \\ Refs.~\citenum{cencek12,przybytek17}}} \\
 &  & 6 & 3766.72 & $\pm${\bf 3.2} & $\pm$7.2 & $\pm$12 & 1.01 \\
 &  & 7 & 3766.04 & $\pm$0.46 & $\pm$1.1 & $\pm${\bf 1.7} & 1.69 \\
\hline
\multirow{2}{*}{\makecell{benzene dimer \\ int. energy (in kcal/mol)}} &
\multirow{2}{*}{\makecell{ MP2 \\ A'V$X$Z, Ref.~\citenum{karton21} }} 
    & 4 & 9.265 & $\pm${\bf 0.43} & $\pm$0.97 & $\pm$1.6 & 0.028 &
\multirow{2}{*}{\makecell{9.293 \\ Ref.~\citenum{karton21}}} \\
 &  & 5 & 9.302 & $\pm${\bf 0.025} & $\pm$0.057 & $\pm$0.094 & 0.009 \\
\hline
\multirow{3}{*}{\makecell{argon dimer \\ int. energy (in cm$^{-1}$)}} &
\multirow{3}{*}{\makecell{CCSD(T) \\ d$\uparrow\downarrow$-disp-XZ+(44332) \\ Ref.~\citenum{patkowski10} }} 
    & 5 & 97.294 & $\pm${\bf 0.76} & $\pm$1.7 & $\pm$2.8 & 0.151 &
\multirow{3}{*}{\makecell{97.445$\pm$0.063 \\ Ref.~\citenum{patkowski10}}} \\
 &  & 6 & 97.515 & $\pm${\bf 0.15} & $\pm$0.33 & $\pm$0.55 & 0.070 \\
  \\
\hline 
\multirow{4}{*}{\makecell{He atom \\ dipole polarizability \\ (in a.u.)}} &
\multirow{4}{*}{\makecell{FCI \\ d$X$Z \\ Refs.~\citenum{lesiuk19,cencek12} }} 
    & 4 & 1.383061 & $\pm${\bf 0.00022} & $\pm$0.00049 & $\pm$0.00082 & 0.000131 &
\multirow{4}{*}{\makecell{1.383192 \\ Ref.~\citenum{pachucki00}}} \\
 &  & 5* & 1.383096 & $\pm$0.000023 & $\pm$0.000052 & $\pm$0.000086 & 0.000097 \\
 &  & 6 & 1.383147 & $\pm$0.000035 & $\pm${\bf 0.000077} & $\pm$0.00013 & 0.000045 \\
 &  & 7 & 1.383170 & $\pm$0.000015 & $\pm${\bf 0.000034} & $\pm$0.000057 & 0.000022 \\
\hline
\multirow{3}{*}{\makecell{H$_2$ molecule \\ dipole polarizability \\ (in a.u.)}} &
\multirow{3}{*}{\makecell{FCI \\ aug-mcc-pVXZ \\ Ref.~\citenum{mielke99} }} 
    & 3 & 6.38944 & $\pm${\bf 0.0068} & $\pm$0.016 & $\pm$0.026 & 0.00212 &
\multirow{3}{*}{\makecell{6.38732  \\ Ref.~\citenum{rychlewski80}}} \\
 &  & 4 & 6.38731 & $\pm${\bf 0.0015} & $\pm$0.0032 & $\pm$0.0053 & 0.00001 \\
 &  & 5 & 6.38772 & $\pm$0.00028 & $\pm${\bf 0.00062} & $\pm$0.0011 & 0.00041 \\
\hline
\multirow{5}{*}{\makecell{Ne atom \\ dipole polarizability \\ (in a.u.$\cdot 10^3$)}} &
\multirow{5}{*}{\makecell{$\Delta$CCSD \\ q-aug-nZP' \\ Ref.~\citenum{hellmann22} }} 
    & 7 & $-$33.431 & $\pm$0.13 & $\pm${\bf 0.28} & $\pm$0.46 & 0.166 &
\multirow{5}{*}{\makecell{$-$33.265$\pm$0.003\\ Ref.~\citenum{hellmann22}}} \\
 &  & 8 & $-$33.357 & $\pm$0.050 & $\pm${\bf 0.12} & $\pm$0.19 & 0.092 \\
 &  & 9 & $-$33.315 & $\pm$0.028 & $\pm${\bf 0.063} & $\pm$0.11 & 0.050 \\
 &  &10 & $-$33.300 & $\pm$0.010 & $\pm$0.023 & $\pm${\bf 0.038} & 0.035 \\
 &  &11 & $-$33.289 & $\pm$0.008 & $\pm$0.017 & $\pm${\bf 0.027} & 0.024 \\
\hline
\multirow{5}{*}{\makecell{Ar atom \\ dipole polarizability \\ (in a.u.)}} &
\multirow{5}{*}{\makecell{$\Delta$CCSD \\ da$X$Z \\ Ref.~\citenum{lesiuk23} }} 
    & 4 & $-$0.3794 & $\pm${\bf 0.14} & $\pm$0.31 & $\pm$0.51 & 0.0152 &
\multirow{5}{*}{\makecell{$-$0.3642 $\pm$ 0.0004 \\ Ref.~\citenum{pachucki04}}} \\
 &  & 5 & $-$0.3536 & $\pm${\bf 0.018} & $\pm$0.039 & $\pm$0.064 & 0.0106 \\
 &  & 6 & $-$0.3620 & $\pm${\bf 0.0056} & $\pm$0.013 & $\pm$0.021 & 0.0022 \\
 &  & 7* & $-$0.3622 &  $\pm$0.0002 & $\pm$0.0003 & $\pm$0.0005 & 0.0020 \\
 &  & 8 & $-$0.3633 & $\pm$0.0008 & $\pm${\bf 0.0016} & $\pm$0.0027 & 0.0009 \\
 \hline\hline
  \multicolumn{9}{l}{$^a$absolute error with respect to the reference value given in the last column}\\

\end{tabular}
}
\end{table*}

For the purpose of further analysis, we call the uncertainty prediction successful at a given confidence level if the true error evaluated against the reference data falls within the determined error bars. Gathering all atoms/molecules, properties and basis set combinations included in Tables~\ref{tab:tests_corr}~and~\ref{tab:tests_prop}, we have 71 distinct sets of data to which the proposed uncertainty prediction procedure was applied. 
Out of that, uncertainty prediction at $1\sigma$ level is found to be successful in about 54\% of cases and $2\sigma$ level in 82\% of cases. We have encountered only two cases where the $3\sigma$ level is unsuccessful (denoted by asterisks in Table~\ref{tab:tests_prop}) and we will discuss these examples in detail further in the text. First, let us put the obtained results into perspective by comparing these percentages with two other popular schemes for attaching an uncertainty to the extrapolated results. The first is the difference between two consecutive extrapolated results, i.e., $|e_X - e_{X-1}|$, while the second is the difference between the extrapolated result and the corresponding result in the largest basis set available, i.e., $|e_X - E_X|$. The first method is successful only in about 25\% of cases considered in Tables~\ref{tab:tests_corr}~and~\ref{tab:tests_prop}, so clearly it is not a reliable indicator of the residual basis set incompleteness error. The second method is successful in most cases considered in Tables~\ref{tab:tests_corr}~and~\ref{tab:tests_prop}, but the error bars determined in this way are usually very broad. Therefore, the use of this approach lead to gross overestimation of the actual error, making it a much less attractive method in practice.

Returning to the examples where error prediction at $3\sigma$ level is not successful, the origin of the problem is traced back to the violation of the fundamental assumption of our method, namely the monotonic decrease of the absolute differences between extrapolated results. Taking the polarizability of argon atom as an example, the extrapolated results in this case are $-$0.3620, $-$0.3622, and $-$0.3633 for $X=6,7,8$. Clearly, the difference between $e_6$ and $e_7$ is smaller here than between $e_7$ and $e_8$, violating the assumptions from Eqs.~(\ref{bound1})~and~(\ref{bound2}). 
One could argue that in such situations the proposed method should not be used at all or a different extrapolation scheme should be applied to eliminate this pathological behavior. 
However, we propose a simple modification of the procedure in such situations: use the difference $|e_{X} - e_{X-2}|$ rather than $|e_X - e_{X-1}|$ to initiate the random walk starting with $e_X$. After this straightforward modification, the result at $1\sigma$ uncertainty level becomes $-$0.3622$\,\pm\,$0.0057 and the true error ($0.0020$) is well within the determined error bars. 
Using the aforementioned procedure with the second problematic case (helium polarizability) we obtain  $1.38310\,\pm\,0.00020$ at $1\sigma$ uncertainty level with the true error equal to $0.00010$.

\begin{figure}
    \centering\hspace{-0.5cm}
    \includegraphics[scale=0.8]{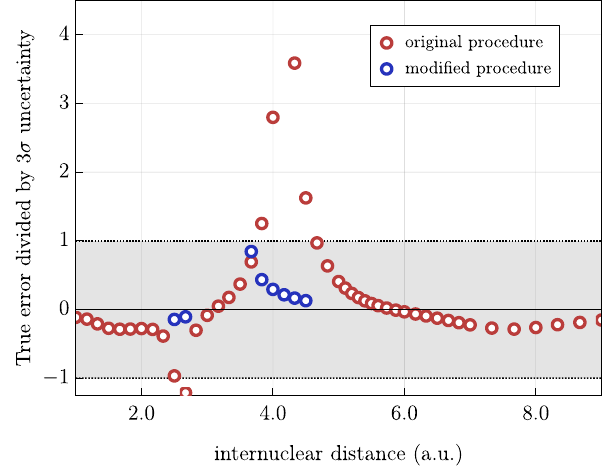}
    \caption{Performance of the original vs.\ modified uncertainty prediction procedure for the interaction energy of the helium dimer (FCI method) as a function of the internuclear distance. Progression of basis sets d$X$Z with $X=5,6,7$ is used to initiate the random walks. On the vertical axis we show the ratio of true error of the extrapolation (with respect to the reference~\cite{cencek12,przybytek17}) and the uncertainty predicted at the $3\sigma$ confidence level. The region where the uncertainty prediction is considered successful (corresponding to the ratio within the interval $[-1,+1]$) is shaded gray. The uncertainties were determined using the original (red points) and modified procedures (blue dots), see text.}
    \label{fig:he2}
\end{figure}

However, the success of the modified procedure in this single case is not sufficient to claim that it performs equally well in general. To address this, we looked for other examples where the fundamental assumption is violated. A handful of them are found in Tables~\ref{tab:tests_corr}~and~\ref{tab:tests_prop}, but deviations from monotonicity of $e_X$ are small and the unmodified procedure predicts the error successfully. However, we encountered significant violations of the fundamental assumption in the interaction energies of helium dimer taken from Refs.~\citenum{cencek12,przybytek17}. For example, for the internuclear distance $R=4.17\,$a.u., the extrapolated results $e_X$ are $176.59$, $178.60$, $178.59$, and $178.30\,$K for $X=4,5,6,7$. Clearly, the middle two numbers are accidentally close to each other and the differences between the extrapolations do not behave monotonically. As illustrated in Fig.~\ref{fig:he2}, this leads to significant underestimation of the uncertainties at $R=4.17\,$a.u. (and, for the same reason, at a handful of neighboring points on the interaction energy curve). When the proposed modification was applied to all points for which non-monotonic behavior was observed, the problem of underestimated uncertainty was solved, see Fig.~\ref{fig:he2}. In the same spirit, the difference $|e_X - E_X|$ can be used as an even more conservative initial bound for the next extrapolation in situations where the value of $e_{X-2}$ is not available.

Finally, we observe that the $1\sigma$ confidence interval performs particularly well when applied to results obtained from three smallest basis sets, $X=2,3,4$. In this case, the method is successful in a significantly larger percentage of cases than one would expect from its confidence level. This may be a consequence of the fact that in smaller basis sets, $X=2$ in particular, other sources of error than lack of higher angular momentum functions remain significant. Insufficient radial saturation, i.e., too small number of functions for angular momenta included in the basis, may be the major contributing factor here. While these secondary sources of error typically converge faster as a function of the basis set size, they are effectively extrapolated according to the $X^{-3}$ rule, leading to their slight overestimation.

The results presented above were based on the extrapolation scheme of Helgaker~\emph{et al.}~\cite{helgaker97,halkier98}. However, other extrapolation techniques are also frequently used in the literature and it is interesting to compare their respective uncertainties predicted by the proposed method. To this end, we selected four distinct two-point extrapolation schemes: (1) $X^{-3}$ method of Helgaker~\emph{et al.}~\cite{helgaker97,halkier98} (same as above), (2) $(X+1/2)^{-4}$ scheme of Martin~\cite{martin96}, (3) method based on Riemann zeta function~\cite{lesiuk19b}, and (4) the scheme proposed by Varandas where the parameter $X$ characterizing the basis set size is adjustable~\cite{varandas07,varandas18,varandas21}. For the purpose of this test, we return to the same systems and basis set combinations as in Table~\ref{tab:tests12} and perform analogous calculations using the aforementioned four extrapolation methods. In the extrapolation scheme (4), the hierarchical numbers for V$X$Z and AV$X$Z basis sets~\cite{varandas21} were used in test case 1 and test case 2, respectively. The results are included in Table~\ref{tab:testfour} at the $1\sigma$, $2\sigma$, and $3\sigma$ confidence levels.

\begin{table}
\caption{\label{tab:testfour}
Comparison of uncertainties assigned to the extrapolated results based on four different extrapolation methods (see text for definitions). All values are given in mHa (with signs reversed for convenience).}
\begin{tabular}{ccc}
\hline\hline
Extrapolation & Test case 1 & Test case 2 \\
\hline
 & \multicolumn{2}{c}{$1\sigma\,$(68.27\%)}  \\
\hline
(1) & 40.8378$\,\pm\,$0.0078 & 154.7$\,\pm\,$2.2 \\
(2) & 40.824$\,\pm\,$0.012 & 154.0$\,\pm\,$2.2 \\
(3) & 40.8455$\,\pm\,$0.0026 & 155.7$\,\pm\,$1.1 \\
(4) & 40.828$\,\pm\,$0.013 & 154.2$\,\pm\,$1.1 \\
\hline
 & \multicolumn{2}{c}{$2\sigma\,$(95.45\%)}  \\
\hline
(1) & 40.838$\,\pm\,$0.018 & 154.7$\,\pm\,$4.8 \\
(2) & 40.824$\,\pm\,$0.027 & 154.0$\,\pm\,$5.0 \\
(3) & 40.8455$\,\pm\,$0.0058 & 155.7$\,\pm\,$2.5 \\
(4) & 40.828$\,\pm\,$0.028 & 154.2$\,\pm\,$2.4 \\
\hline
 & \multicolumn{2}{c}{$3\sigma\,$(99.73\%)}  \\
\hline
(1) & 40.838$\,\pm\,$0.029 & 154.7$\,\pm\,$7.9 \\
(2) & 40.824$\,\pm\,$0.045 & 154.0$\,\pm\,$8.3 \\
(3) & 40.8455$\,\pm\,$0.0096 & 155.7$\,\pm\,$4.1 \\
(4) & 40.828$\,\pm\,$0.047 & 154.2$\,\pm\,$3.9 \\
\hline
reference & 40.8463 & 156.287 \\
\hline\hline
\end{tabular}
\end{table}
 
The data reported in Table~\ref{tab:testfour} leads to the conclusion that all extrapolation schemes give consistent results, if their respective uncertainties are taken into account. Even if we consider a pair of extrapolation schemes which differ the most from each other ($40.824$ vs. $40.846$ for test case 1; $154.0$ vs. $155.7$ for test case 2), the differences are smaller than the sum of their uncertainties at the $2\sigma$ level, $0.033$ and $7.5$, respectively. Simultaneously, for all data points in Table~\ref{tab:testfour} the differences between the extrapolated results and the corresponding reference values are smaller than the uncertainty at the $2\sigma$ level.

To sum up, we have introduced a method of estimating the uncertainty of a result obtained through extrapolation to the complete basis set limit. The method is based on an ensemble of random walks which simulate possible extrapolation outcomes that could have been obtained if results from larger basis sets had been available. The ensemble of independent random walks is then analyzed statistically, enabling uncertainty prediction at a given confidence level. The method is free of empiricism and can be used in conjunction with any extrapolation scheme. Numerical tests performed in this work show that the proposed method is successful in predicting the extrapolation error, leading to error bars which are tight yet conservative at the same time. While the extrapolation error is the natural target for the proposed procedure, it is possible that similar ideas can be used to determine uncertainties due to other sources of error in quantum-chemical calculations and beyond.

A {\sc Python} implementation of the proposed procedure is available open-source on GitHub ({\url{https://github.com/lesiukmichal/extrapolation-random-walk}}) ~\cite{github}.

\begin{acknowledgement}
M.L. was supported by the National Science Centre, Poland, under research project 2022/47/D/ST4/01834. We gratefully acknowledge Poland’s high-performance Infrastructure PLGrid (HPC Centers: ACK Cyfronet AGH, PCSS, CI TASK, WCSS) for providing computer facilities and support within computational grants PLG/2023/016599 and PLG/2024/017370. 
\end{acknowledgement}

\bibliography{extra-error}

\providecommand{\latin}[1]{#1}
\makeatletter
\providecommand{\doi}
  {\begingroup\let\do\@makeother\dospecials
  \catcode`\{=1 \catcode`\}=2 \doi@aux}
\providecommand{\doi@aux}[1]{\endgroup\texttt{#1}}
\makeatother
\providecommand*\mcitethebibliography{\thebibliography}
\csname @ifundefined\endcsname{endmcitethebibliography}
  {\let\endmcitethebibliography\endthebibliography}{}
\begin{mcitethebibliography}{77}
\providecommand*\natexlab[1]{#1}
\providecommand*\mciteSetBstSublistMode[1]{}
\providecommand*\mciteSetBstMaxWidthForm[2]{}
\providecommand*\mciteBstWouldAddEndPuncttrue
  {\def\EndOfBibitem{\unskip.}}
\providecommand*\mciteBstWouldAddEndPunctfalse
  {\let\EndOfBibitem\relax}
\providecommand*\mciteSetBstMidEndSepPunct[3]{}
\providecommand*\mciteSetBstSublistLabelBeginEnd[3]{}
\providecommand*\EndOfBibitem{}
\mciteSetBstSublistMode{f}
\mciteSetBstMaxWidthForm{subitem}{(\alph{mcitesubitemcount})}
\mciteSetBstSublistLabelBeginEnd
  {\mcitemaxwidthsubitemform\space}
  {\relax}
  {\relax}

\bibitem[Tomza \latin{et~al.}(2019)Tomza, Jachymski, Gerritsma, Negretti,
  Calarco, Idziaszek, and Julienne]{tomza19}
Tomza,~M.; Jachymski,~K.; Gerritsma,~R.; Negretti,~A.; Calarco,~T.;
  Idziaszek,~Z.; Julienne,~P.~S. Cold hybrid ion-atom systems. \emph{Rev. Mod.
  Phys.} \textbf{2019}, \emph{91}, 035001\relax
\mciteBstWouldAddEndPuncttrue
\mciteSetBstMidEndSepPunct{\mcitedefaultmidpunct}
{\mcitedefaultendpunct}{\mcitedefaultseppunct}\relax
\EndOfBibitem
\bibitem[Gronowski \latin{et~al.}(2020)Gronowski, Koza, and Tomza]{gronowski20}
Gronowski,~M.; Koza,~A.~M.; Tomza,~M. Ab initio properties of the NaLi molecule
  in the $a^3\Sigma^+$ electronic state. \emph{Phys. Rev. A} \textbf{2020},
  \emph{102}, 020801\relax
\mciteBstWouldAddEndPuncttrue
\mciteSetBstMidEndSepPunct{\mcitedefaultmidpunct}
{\mcitedefaultendpunct}{\mcitedefaultseppunct}\relax
\EndOfBibitem
\bibitem[{\'S}mia{\l}kowski and Tomza(2021){\'S}mia{\l}kowski, and
  Tomza]{smialkowski21}
{\'S}mia{\l}kowski,~M.; Tomza,~M. Highly polar molecules consisting of a copper
  or silver atom interacting with an alkali-metal or alkaline-earth-metal atom.
  \emph{Phys. Rev. A} \textbf{2021}, \emph{103}, 022802\relax
\mciteBstWouldAddEndPuncttrue
\mciteSetBstMidEndSepPunct{\mcitedefaultmidpunct}
{\mcitedefaultendpunct}{\mcitedefaultseppunct}\relax
\EndOfBibitem
\bibitem[Ladjimi and Tomza(2023)Ladjimi, and Tomza]{ladjimi23}
Ladjimi,~H.; Tomza,~M. Chemical reactions of ultracold alkaline-earth-metal
  diatomic molecules. \emph{Phys. Rev. A} \textbf{2023}, \emph{108},
  L021302\relax
\mciteBstWouldAddEndPuncttrue
\mciteSetBstMidEndSepPunct{\mcitedefaultmidpunct}
{\mcitedefaultendpunct}{\mcitedefaultseppunct}\relax
\EndOfBibitem
\bibitem[Finelli \latin{et~al.}(2024)Finelli, Ciamei, Restivo, Schemmer, Cosco,
  Inguscio, Trenkwalder, Zaremba-Kopczyk, Gronowski, Tomza, and
  Zaccanti]{finelli24}
Finelli,~S.; Ciamei,~A.; Restivo,~B.; Schemmer,~M.; Cosco,~A.; Inguscio,~M.;
  Trenkwalder,~A.; Zaremba-Kopczyk,~K.; Gronowski,~M.; Tomza,~M. \latin{et~al.}
   Ultracold LiCr: A New Pathway to Quantum Gases of Paramagnetic Polar
  Molecules. \emph{PRX Quantum} \textbf{2024}, \emph{5}, 020358\relax
\mciteBstWouldAddEndPuncttrue
\mciteSetBstMidEndSepPunct{\mcitedefaultmidpunct}
{\mcitedefaultendpunct}{\mcitedefaultseppunct}\relax
\EndOfBibitem
\bibitem[Lesiuk \latin{et~al.}(2020)Lesiuk, Przybytek, and
  Jeziorski]{lesiuk20b}
Lesiuk,~M.; Przybytek,~M.; Jeziorski,~B. Theoretical determination of
  polarizability and magnetic susceptibility of neon. \emph{Phys. Rev. A}
  \textbf{2020}, \emph{102}, 052816\relax
\mciteBstWouldAddEndPuncttrue
\mciteSetBstMidEndSepPunct{\mcitedefaultmidpunct}
{\mcitedefaultendpunct}{\mcitedefaultseppunct}\relax
\EndOfBibitem
\bibitem[Hellmann(2022)]{hellmann22}
Hellmann,~R. Ab initio determination of the polarizability of neon. \emph{Phys.
  Rev. A} \textbf{2022}, \emph{105}, 022809\relax
\mciteBstWouldAddEndPuncttrue
\mciteSetBstMidEndSepPunct{\mcitedefaultmidpunct}
{\mcitedefaultendpunct}{\mcitedefaultseppunct}\relax
\EndOfBibitem
\bibitem[Lang \latin{et~al.}(2023)Lang, Garberoglio, Przybytek, Jeziorska, and
  Jeziorski]{lang23}
Lang,~J.; Garberoglio,~G.; Przybytek,~M.; Jeziorska,~M.; Jeziorski,~B.
  Three-body potential and third virial coefficients for helium including
  relativistic and nuclear-motion effects. \emph{Phys. Chem. Chem. Phys.}
  \textbf{2023}, \emph{25}, 23395--23416\relax
\mciteBstWouldAddEndPuncttrue
\mciteSetBstMidEndSepPunct{\mcitedefaultmidpunct}
{\mcitedefaultendpunct}{\mcitedefaultseppunct}\relax
\EndOfBibitem
\bibitem[Lang \latin{et~al.}(2023)Lang, Przybytek, Lesiuk, and
  Jeziorski]{lang2023collision}
Lang,~J.; Przybytek,~M.; Lesiuk,~M.; Jeziorski,~B. Collision-induced three-body
  polarizability of helium. \emph{J. Chem. Phys.} \textbf{2023}, \emph{158},
  114303\relax
\mciteBstWouldAddEndPuncttrue
\mciteSetBstMidEndSepPunct{\mcitedefaultmidpunct}
{\mcitedefaultendpunct}{\mcitedefaultseppunct}\relax
\EndOfBibitem
\bibitem[Lesiuk and Jeziorski(2023)Lesiuk, and Jeziorski]{lesiuk23}
Lesiuk,~M.; Jeziorski,~B. First-principles calculation of the
  frequency-dependent dipole polarizability of argon. \emph{Phys. Rev. A}
  \textbf{2023}, \emph{107}, 042805\relax
\mciteBstWouldAddEndPuncttrue
\mciteSetBstMidEndSepPunct{\mcitedefaultmidpunct}
{\mcitedefaultendpunct}{\mcitedefaultseppunct}\relax
\EndOfBibitem
\bibitem[Garberoglio \latin{et~al.}(2023)Garberoglio, Gaiser, Gavioso, Harvey,
  Hellmann, Jeziorski, Meier, Moldover, Pitre, Szalewicz, \latin{et~al.}
  others]{garberoglio23}
Garberoglio,~G.; Gaiser,~C.; Gavioso,~R.~M.; Harvey,~A.~H.; Hellmann,~R.;
  Jeziorski,~B.; Meier,~K.; Moldover,~M.~R.; Pitre,~L.; Szalewicz,~K.
  \latin{et~al.}  Ab initio calculation of fluid properties for precision
  metrology. \emph{J. Phys. Chem. Ref. Data} \textbf{2023}, \emph{52},
  031502\relax
\mciteBstWouldAddEndPuncttrue
\mciteSetBstMidEndSepPunct{\mcitedefaultmidpunct}
{\mcitedefaultendpunct}{\mcitedefaultseppunct}\relax
\EndOfBibitem
\bibitem[Lesiuk \latin{et~al.}(2019)Lesiuk, Przybytek, Balcerzak, Musia{\l},
  and Moszynski]{lesiuk19}
Lesiuk,~M.; Przybytek,~M.; Balcerzak,~J.~G.; Musia{\l},~M.; Moszynski,~R. Ab
  initio potential energy curve for the ground state of beryllium dimer.
  \emph{J. Chem. Theory Comput.} \textbf{2019}, \emph{15}, 2470--2480\relax
\mciteBstWouldAddEndPuncttrue
\mciteSetBstMidEndSepPunct{\mcitedefaultmidpunct}
{\mcitedefaultendpunct}{\mcitedefaultseppunct}\relax
\EndOfBibitem
\bibitem[Lesiuk \latin{et~al.}(2020)Lesiuk, Musia{\l}, and Moszynski]{lesiuk20}
Lesiuk,~M.; Musia{\l},~M.; Moszynski,~R. Potential-energy curve for the
  $a^3\Sigma_u^+$ state of a lithium dimer with Slater-type orbitals.
  \emph{Phys. Rev. A} \textbf{2020}, \emph{102}, 062806\relax
\mciteBstWouldAddEndPuncttrue
\mciteSetBstMidEndSepPunct{\mcitedefaultmidpunct}
{\mcitedefaultendpunct}{\mcitedefaultseppunct}\relax
\EndOfBibitem
\bibitem[G{\c{e}}bala \latin{et~al.}(2023)G{\c{e}}bala, Przybytek, Gronowski,
  and Tomza]{gebala23}
G{\c{e}}bala,~J.; Przybytek,~M.; Gronowski,~M.; Tomza,~M. Ab initio
  potential-energy curves, scattering lengths, and rovibrational levels of the
  He$_2^+$ molecular ion in excited electronic states. \emph{Phys. Rev. A}
  \textbf{2023}, \emph{108}, 052821\relax
\mciteBstWouldAddEndPuncttrue
\mciteSetBstMidEndSepPunct{\mcitedefaultmidpunct}
{\mcitedefaultendpunct}{\mcitedefaultseppunct}\relax
\EndOfBibitem
\bibitem[Landau \latin{et~al.}(2023)Landau, Eduardus, Behar, Wallach,
  Pa{\v{s}}teka, Faraji, Borschevsky, and Shagam]{landau23}
Landau,~A.; Eduardus; Behar,~D.; Wallach,~E.~R.; Pa{\v{s}}teka,~L.~F.;
  Faraji,~S.; Borschevsky,~A.; Shagam,~Y. Chiral molecule candidates for
  trapped ion spectroscopy by ab initio calculations: From state preparation to
  parity violation. \emph{J. Chem. Phys.} \textbf{2023}, \emph{159},
  114307\relax
\mciteBstWouldAddEndPuncttrue
\mciteSetBstMidEndSepPunct{\mcitedefaultmidpunct}
{\mcitedefaultendpunct}{\mcitedefaultseppunct}\relax
\EndOfBibitem
\bibitem[ISOLDE-Collaboration \latin{et~al.}(2025)ISOLDE-Collaboration,
  \latin{et~al.} others]{isolde25}
ISOLDE-Collaboration; others Pinning down electron correlations in RaF via
  spectroscopy of excited states and high-accuracy relativistic quantum
  chemistry. \emph{Nat. Commun.} \textbf{2025}, accepted, in press\relax
\mciteBstWouldAddEndPuncttrue
\mciteSetBstMidEndSepPunct{\mcitedefaultmidpunct}
{\mcitedefaultendpunct}{\mcitedefaultseppunct}\relax
\EndOfBibitem
\bibitem[Safronova \latin{et~al.}(2018)Safronova, Budker, DeMille, Kimball,
  Derevianko, and Clark]{safronova18}
Safronova,~M.; Budker,~D.; DeMille,~D.; Kimball,~D. F.~J.; Derevianko,~A.;
  Clark,~C.~W. Search for new physics with atoms and molecules. \emph{Rev. Mod.
  Phys.} \textbf{2018}, \emph{90}, 025008\relax
\mciteBstWouldAddEndPuncttrue
\mciteSetBstMidEndSepPunct{\mcitedefaultmidpunct}
{\mcitedefaultendpunct}{\mcitedefaultseppunct}\relax
\EndOfBibitem
\bibitem[Gaul \latin{et~al.}(2019)Gaul, Marquardt, Isaev, and Berger]{gaul19}
Gaul,~K.; Marquardt,~S.; Isaev,~T.; Berger,~R. Systematic study of relativistic
  and chemical enhancements of P, T-odd effects in polar diatomic radicals.
  \emph{Phys. Rev. A} \textbf{2019}, \emph{99}, 032509\relax
\mciteBstWouldAddEndPuncttrue
\mciteSetBstMidEndSepPunct{\mcitedefaultmidpunct}
{\mcitedefaultendpunct}{\mcitedefaultseppunct}\relax
\EndOfBibitem
\bibitem[Tomza(2021)]{tomza21}
Tomza,~M. Interaction potentials, electric moments, polarizabilities, and
  chemical reactions of YbCu, YbAg, and YbAu molecules. \emph{New J. Phys.}
  \textbf{2021}, \emph{23}, 085003\relax
\mciteBstWouldAddEndPuncttrue
\mciteSetBstMidEndSepPunct{\mcitedefaultmidpunct}
{\mcitedefaultendpunct}{\mcitedefaultseppunct}\relax
\EndOfBibitem
\bibitem[Marc \latin{et~al.}(2023)Marc, Hubert, and Fleig]{marc23}
Marc,~A.; Hubert,~M.; Fleig,~T. Candidate molecules for next-generation
  searches of hadronic charge-parity violation. \emph{Phys. Rev. A}
  \textbf{2023}, \emph{108}, 062815\relax
\mciteBstWouldAddEndPuncttrue
\mciteSetBstMidEndSepPunct{\mcitedefaultmidpunct}
{\mcitedefaultendpunct}{\mcitedefaultseppunct}\relax
\EndOfBibitem
\bibitem[Chamorro \latin{et~al.}(2024)Chamorro, Flambaum, Garcia~Ruiz,
  Borschevsky, and Pa{\v{s}}teka]{chamorro24}
Chamorro,~Y.; Flambaum,~V.~V.; Garcia~Ruiz,~R.~F.; Borschevsky,~A.;
  Pa{\v{s}}teka,~L.~F. Enhanced parity and time-reversal-symmetry violation in
  diatomic molecules: LaO, LaS, and LuO. \emph{Phys. Rev. A} \textbf{2024},
  \emph{110}, 042806\relax
\mciteBstWouldAddEndPuncttrue
\mciteSetBstMidEndSepPunct{\mcitedefaultmidpunct}
{\mcitedefaultendpunct}{\mcitedefaultseppunct}\relax
\EndOfBibitem
\bibitem[Kato(1957)]{kato57}
Kato,~T. {On the eigenfunctions of many-particle systems in quantum mechanics}.
  \emph{Commun. Pure Appl. Math.} \textbf{1957}, \emph{10}, 151--177\relax
\mciteBstWouldAddEndPuncttrue
\mciteSetBstMidEndSepPunct{\mcitedefaultmidpunct}
{\mcitedefaultendpunct}{\mcitedefaultseppunct}\relax
\EndOfBibitem
\bibitem[Helgaker \latin{et~al.}(2000)Helgaker, Jørgensen, and
  Olsen]{helgaker00}
Helgaker,~T.; Jørgensen,~P.; Olsen,~J. \emph{{Molecular Electronic-Structure
  Theory}}; {Wiley}, 2000\relax
\mciteBstWouldAddEndPuncttrue
\mciteSetBstMidEndSepPunct{\mcitedefaultmidpunct}
{\mcitedefaultendpunct}{\mcitedefaultseppunct}\relax
\EndOfBibitem
\bibitem[H\"attig \latin{et~al.}(2012)H\"attig, Klopper, K\"ohn, and
  Tew]{haettig12}
H\"attig,~C.; Klopper,~W.; K\"ohn,~A.; Tew,~D.~P. {Explicitly Correlated
  Electrons in Molecules}. \emph{Chem. Rev.} \textbf{2012}, \emph{112},
  4--74\relax
\mciteBstWouldAddEndPuncttrue
\mciteSetBstMidEndSepPunct{\mcitedefaultmidpunct}
{\mcitedefaultendpunct}{\mcitedefaultseppunct}\relax
\EndOfBibitem
\bibitem[Kong \latin{et~al.}(2012)Kong, Bischoff, and Valeev]{kong12}
Kong,~L.; Bischoff,~F.~A.; Valeev,~E.~F. {Explicitly Correlated R12/F12 Methods
  for Electronic Structure}. \emph{Chem. Rev.} \textbf{2012}, \emph{112},
  75--107\relax
\mciteBstWouldAddEndPuncttrue
\mciteSetBstMidEndSepPunct{\mcitedefaultmidpunct}
{\mcitedefaultendpunct}{\mcitedefaultseppunct}\relax
\EndOfBibitem
\bibitem[Ten-no(2012)]{tenno12}
Ten-no,~S. {Explicitly correlated wave functions: summary and perspective}.
  \emph{Theor. Chem. Acc.} \textbf{2012}, \emph{131}, 1070\relax
\mciteBstWouldAddEndPuncttrue
\mciteSetBstMidEndSepPunct{\mcitedefaultmidpunct}
{\mcitedefaultendpunct}{\mcitedefaultseppunct}\relax
\EndOfBibitem
\bibitem[Boys \latin{et~al.}(1969)Boys, Handy, and Linnett]{boiii69}
Boys,~S.~F.; Handy,~N.~C.; Linnett,~J.~W. {The determination of energies and
  wavefunctions with full electronic correlation}. \emph{Proc. R. Soc. Lond.}
  \textbf{1969}, \emph{310}, 43--61\relax
\mciteBstWouldAddEndPuncttrue
\mciteSetBstMidEndSepPunct{\mcitedefaultmidpunct}
{\mcitedefaultendpunct}{\mcitedefaultseppunct}\relax
\EndOfBibitem
\bibitem[Loos \latin{et~al.}(2019)Loos, Pradines, Scemama, Toulouse, and
  Giner]{loos19a}
Loos,~P.-F.; Pradines,~B.; Scemama,~A.; Toulouse,~J.; Giner,~E. A density-based
  basis-set correction for wave function theory. \emph{J. Phys. Chem. Lett.}
  \textbf{2019}, \emph{10}, 2931--2937\relax
\mciteBstWouldAddEndPuncttrue
\mciteSetBstMidEndSepPunct{\mcitedefaultmidpunct}
{\mcitedefaultendpunct}{\mcitedefaultseppunct}\relax
\EndOfBibitem
\bibitem[Loos \latin{et~al.}(2019)Loos, Pradines, Scemama, Giner, and
  Toulouse]{loos19b}
Loos,~P.-F.; Pradines,~B.; Scemama,~A.; Giner,~E.; Toulouse,~J. Density-based
  basis-set incompleteness correction for GW methods. \emph{J. Chem. Theory
  Comput.} \textbf{2019}, \emph{16}, 1018--1028\relax
\mciteBstWouldAddEndPuncttrue
\mciteSetBstMidEndSepPunct{\mcitedefaultmidpunct}
{\mcitedefaultendpunct}{\mcitedefaultseppunct}\relax
\EndOfBibitem
\bibitem[Giner \latin{et~al.}(2020)Giner, Scemama, Loos, and Toulouse]{giner20}
Giner,~E.; Scemama,~A.; Loos,~P.-F.; Toulouse,~J. A basis-set error correction
  based on density-functional theory for strongly correlated molecular systems.
  \emph{J. Chem. Phys.} \textbf{2020}, \emph{152}, 174104\relax
\mciteBstWouldAddEndPuncttrue
\mciteSetBstMidEndSepPunct{\mcitedefaultmidpunct}
{\mcitedefaultendpunct}{\mcitedefaultseppunct}\relax
\EndOfBibitem
\bibitem[Mester and K{\'a}llay(2025)Mester, and K{\'a}llay]{mester25}
Mester,~D.; K{\'a}llay,~M. Higher-order coupled-cluster calculations with
  basis-set corrections. \emph{Chem. Phys. Lett.} \textbf{2025}, \emph{861},
  141780\relax
\mciteBstWouldAddEndPuncttrue
\mciteSetBstMidEndSepPunct{\mcitedefaultmidpunct}
{\mcitedefaultendpunct}{\mcitedefaultseppunct}\relax
\EndOfBibitem
\bibitem[Klopper(2001)]{klopper01}
Klopper,~W. Highly accurate coupled-cluster singlet and triplet pair energies
  from explicitly correlated calculations in comparison with extrapolation
  techniques. \emph{Mol. Phys.} \textbf{2001}, \emph{99}, 481--507\relax
\mciteBstWouldAddEndPuncttrue
\mciteSetBstMidEndSepPunct{\mcitedefaultmidpunct}
{\mcitedefaultendpunct}{\mcitedefaultseppunct}\relax
\EndOfBibitem
\bibitem[Varandas(2007)]{varandas07}
Varandas,~A. J.~C. Extrapolating to the one-electron basis-set limit in
  electronic structure calculations. \emph{J. Chem. Phys.} \textbf{2007},
  \emph{126}, 244105\relax
\mciteBstWouldAddEndPuncttrue
\mciteSetBstMidEndSepPunct{\mcitedefaultmidpunct}
{\mcitedefaultendpunct}{\mcitedefaultseppunct}\relax
\EndOfBibitem
\bibitem[Feller \latin{et~al.}(2011)Feller, Peterson, and Grant~Hill]{feller11}
Feller,~D.; Peterson,~K.~A.; Grant~Hill,~J. On the effectiveness of CCSD (T)
  complete basis set extrapolations for atomization energies. \emph{J. Chem.
  Phys.} \textbf{2011}, \emph{135}, 044102\relax
\mciteBstWouldAddEndPuncttrue
\mciteSetBstMidEndSepPunct{\mcitedefaultmidpunct}
{\mcitedefaultendpunct}{\mcitedefaultseppunct}\relax
\EndOfBibitem
\bibitem[Varandas(2018)]{varandas18}
Varandas,~A. J.~C. Straightening the hierarchical staircase for basis set
  extrapolations: A low-cost approach to high-accuracy computational chemistry.
  \emph{Annu. Rev. Phys. Chem.} \textbf{2018}, \emph{69}, 177--203\relax
\mciteBstWouldAddEndPuncttrue
\mciteSetBstMidEndSepPunct{\mcitedefaultmidpunct}
{\mcitedefaultendpunct}{\mcitedefaultseppunct}\relax
\EndOfBibitem
\bibitem[Guo \latin{et~al.}(2024)Guo, Pa{\v{s}}teka, Nagame, Sato, Eliav,
  Reitsma, and Borschevsky]{guo24}
Guo,~Y.; Pa{\v{s}}teka,~L.~F.; Nagame,~Y.; Sato,~T.~K.; Eliav,~E.;
  Reitsma,~M.~L.; Borschevsky,~A. Relativistic coupled-cluster calculations of
  the electron affinity and ionization potentials of lawrencium. \emph{Phys.
  Rev. A} \textbf{2024}, \emph{110}, 022817\relax
\mciteBstWouldAddEndPuncttrue
\mciteSetBstMidEndSepPunct{\mcitedefaultmidpunct}
{\mcitedefaultendpunct}{\mcitedefaultseppunct}\relax
\EndOfBibitem
\bibitem[Lang \latin{et~al.}(2024)Lang, Przybytek, and Lesiuk]{lang24}
Lang,~J.; Przybytek,~M.; Lesiuk,~M. Thermophysical properties of argon gas from
  improved two-body interaction potential. \emph{Phys. Rev. A} \textbf{2024},
  \emph{109}, 052803\relax
\mciteBstWouldAddEndPuncttrue
\mciteSetBstMidEndSepPunct{\mcitedefaultmidpunct}
{\mcitedefaultendpunct}{\mcitedefaultseppunct}\relax
\EndOfBibitem
\bibitem[Not()]{Note-1}
While the symbol $E$ is used to denote the quantity of interest throughout the
  paper, it does not imply that we are referring only to energy.\relax
\mciteBstWouldAddEndPunctfalse
\mciteSetBstMidEndSepPunct{\mcitedefaultmidpunct}
{}{\mcitedefaultseppunct}\relax
\EndOfBibitem
\bibitem[Dunning~Jr(1989)]{dunning89}
Dunning~Jr,~T.~H. Gaussian basis sets for use in correlated molecular
  calculations. I. The atoms boron through neon and hydrogen. \emph{J. Chem.
  Phys.} \textbf{1989}, \emph{90}, 1007--1023\relax
\mciteBstWouldAddEndPuncttrue
\mciteSetBstMidEndSepPunct{\mcitedefaultmidpunct}
{\mcitedefaultendpunct}{\mcitedefaultseppunct}\relax
\EndOfBibitem
\bibitem[Carroll \latin{et~al.}(1979)Carroll, Silverstone, and
  Metzger]{carroll79}
Carroll,~D.~P.; Silverstone,~H.~J.; Metzger,~R.~M. Piecewise polynomial
  configuration interaction natural orbital study of $1s^2$ helium. \emph{J.
  Chem. Phys.} \textbf{1979}, \emph{71}, 4142--4163\relax
\mciteBstWouldAddEndPuncttrue
\mciteSetBstMidEndSepPunct{\mcitedefaultmidpunct}
{\mcitedefaultendpunct}{\mcitedefaultseppunct}\relax
\EndOfBibitem
\bibitem[Hill(1985)]{hill85}
Hill,~R.~N. Rates of convergence and error estimation formulas for the
  Rayleigh--Ritz variational method. \emph{J. Chem. Phys.} \textbf{1985},
  \emph{83}, 1173--1196\relax
\mciteBstWouldAddEndPuncttrue
\mciteSetBstMidEndSepPunct{\mcitedefaultmidpunct}
{\mcitedefaultendpunct}{\mcitedefaultseppunct}\relax
\EndOfBibitem
\bibitem[Kutzelnigg and Morgan~III(1992)Kutzelnigg, and
  Morgan~III]{kutzelnigg92}
Kutzelnigg,~W.; Morgan~III,~J.~D. Rates of convergence of the partial-wave
  expansions of atomic correlation energies. \emph{J. Chem. Phys.}
  \textbf{1992}, \emph{96}, 4484--4508\relax
\mciteBstWouldAddEndPuncttrue
\mciteSetBstMidEndSepPunct{\mcitedefaultmidpunct}
{\mcitedefaultendpunct}{\mcitedefaultseppunct}\relax
\EndOfBibitem
\bibitem[Kutzelnigg(2008)]{kutzelnigg08}
Kutzelnigg,~W. Relativistic corrections to the partial wave expansion of
  two-electron atoms. \emph{Int. J. Quantum Chem.} \textbf{2008}, \emph{108},
  2280--2290\relax
\mciteBstWouldAddEndPuncttrue
\mciteSetBstMidEndSepPunct{\mcitedefaultmidpunct}
{\mcitedefaultendpunct}{\mcitedefaultseppunct}\relax
\EndOfBibitem
\bibitem[Helgaker \latin{et~al.}(1997)Helgaker, Klopper, Koch, and
  Noga]{helgaker97}
Helgaker,~T.; Klopper,~W.; Koch,~H.; Noga,~J. Basis-set convergence of
  correlated calculations on water. \emph{J. Chem. Phys.} \textbf{1997},
  \emph{106}, 9639--9646\relax
\mciteBstWouldAddEndPuncttrue
\mciteSetBstMidEndSepPunct{\mcitedefaultmidpunct}
{\mcitedefaultendpunct}{\mcitedefaultseppunct}\relax
\EndOfBibitem
\bibitem[Halkier \latin{et~al.}(1998)Halkier, Helgaker, J{\o}rgensen, Klopper,
  Koch, Olsen, and Wilson]{halkier98}
Halkier,~A.; Helgaker,~T.; J{\o}rgensen,~P.; Klopper,~W.; Koch,~H.; Olsen,~J.;
  Wilson,~A.~K. Basis-set convergence in correlated calculations on Ne, N$_2$,
  and H$_2$O. \emph{Chem. Phys. Lett.} \textbf{1998}, \emph{286},
  243--252\relax
\mciteBstWouldAddEndPuncttrue
\mciteSetBstMidEndSepPunct{\mcitedefaultmidpunct}
{\mcitedefaultendpunct}{\mcitedefaultseppunct}\relax
\EndOfBibitem
\bibitem[Mielke \latin{et~al.}(1999)Mielke, Garrett, and Peterson]{mielke99}
Mielke,~S.~L.; Garrett,~B.~C.; Peterson,~K.~A. The utility of many-body
  decompositions for the accurate basis set extrapolation of ab initio data.
  \emph{J. Chem. Phys.} \textbf{1999}, \emph{111}, 3806--3811\relax
\mciteBstWouldAddEndPuncttrue
\mciteSetBstMidEndSepPunct{\mcitedefaultmidpunct}
{\mcitedefaultendpunct}{\mcitedefaultseppunct}\relax
\EndOfBibitem
\bibitem[Pachucki(2010)]{pachucki10}
Pachucki,~K. Born-Oppenheimer potential for H$_2$. \emph{Phys. Rev. A}
  \textbf{2010}, \emph{82}, 032509\relax
\mciteBstWouldAddEndPuncttrue
\mciteSetBstMidEndSepPunct{\mcitedefaultmidpunct}
{\mcitedefaultendpunct}{\mcitedefaultseppunct}\relax
\EndOfBibitem
\bibitem[Mitin(2000)]{mitin00}
Mitin,~A.~V. Exact solution of the Hartree-Fock equation for the H 2 molecule
  in the linear-combination-of-atomic-orbitals approximation. \emph{Phys. Rev.
  A} \textbf{2000}, \emph{62}, 010501\relax
\mciteBstWouldAddEndPuncttrue
\mciteSetBstMidEndSepPunct{\mcitedefaultmidpunct}
{\mcitedefaultendpunct}{\mcitedefaultseppunct}\relax
\EndOfBibitem
\bibitem[Kendall \latin{et~al.}(1992)Kendall, Dunning~Jr, and
  Harrison]{kendall92}
Kendall,~R.~A.; Dunning~Jr,~T.~H.; Harrison,~R.~J. Electron affinities of the
  first-row atoms revisited. Systematic basis sets and wave functions. \emph{J.
  Chem. Phys.} \textbf{1992}, \emph{96}, 6796--6806\relax
\mciteBstWouldAddEndPuncttrue
\mciteSetBstMidEndSepPunct{\mcitedefaultmidpunct}
{\mcitedefaultendpunct}{\mcitedefaultseppunct}\relax
\EndOfBibitem
\bibitem[Strasburger(2019)]{strasburger19}
Strasburger,~K. Explicitly correlated wave functions of the ground state and
  the lowest quintuplet state of the carbon atom. \emph{Phys. Rev. A}
  \textbf{2019}, \emph{99}, 052512\relax
\mciteBstWouldAddEndPuncttrue
\mciteSetBstMidEndSepPunct{\mcitedefaultmidpunct}
{\mcitedefaultendpunct}{\mcitedefaultseppunct}\relax
\EndOfBibitem
\bibitem[Bunge \latin{et~al.}(1993)Bunge, Barrientos, and Bunge]{bunge93}
Bunge,~C.~F.; Barrientos,~J.~A.; Bunge,~A.~V. Roothaan-Hartree-Fock
  ground-state atomic wave functions: Slater-type orbital expansions and
  expectation values for $Z=2-54$. \emph{At. Data Nucl. Data Tables}
  \textbf{1993}, \emph{53}, 113--162\relax
\mciteBstWouldAddEndPuncttrue
\mciteSetBstMidEndSepPunct{\mcitedefaultmidpunct}
{\mcitedefaultendpunct}{\mcitedefaultseppunct}\relax
\EndOfBibitem
\bibitem[Lesiuk and Jeziorski(2019)Lesiuk, and Jeziorski]{lesiuk19b}
Lesiuk,~M.; Jeziorski,~B. Complete basis set extrapolation of electronic
  correlation energies using the Riemann zeta function. \emph{J. Chem. Theory
  Comput.} \textbf{2019}, \emph{15}, 5398--5403\relax
\mciteBstWouldAddEndPuncttrue
\mciteSetBstMidEndSepPunct{\mcitedefaultmidpunct}
{\mcitedefaultendpunct}{\mcitedefaultseppunct}\relax
\EndOfBibitem
\bibitem[Szalewicz and Jeziorski(2010)Szalewicz, and Jeziorski]{szalewicz10}
Szalewicz,~K.; Jeziorski,~B. Explicitly-correlated Gaussian geminals in
  electronic structure calculations. \emph{Mol. Phys.} \textbf{2010},
  \emph{108}, 3091--3103\relax
\mciteBstWouldAddEndPuncttrue
\mciteSetBstMidEndSepPunct{\mcitedefaultmidpunct}
{\mcitedefaultendpunct}{\mcitedefaultseppunct}\relax
\EndOfBibitem
\bibitem[Mitroy \latin{et~al.}(2013)Mitroy, Bubin, Horiuchi, Suzuki, Adamowicz,
  Cencek, Szalewicz, Komasa, Blume, and Varga]{mitroy13}
Mitroy,~J.; Bubin,~S.; Horiuchi,~W.; Suzuki,~Y.; Adamowicz,~L.; Cencek,~W.;
  Szalewicz,~K.; Komasa,~J.; Blume,~D.; Varga,~K. Theory and application of
  explicitly correlated Gaussians. \emph{Rev. Mod. Phys.} \textbf{2013},
  \emph{85}, 693--749\relax
\mciteBstWouldAddEndPuncttrue
\mciteSetBstMidEndSepPunct{\mcitedefaultmidpunct}
{\mcitedefaultendpunct}{\mcitedefaultseppunct}\relax
\EndOfBibitem
\bibitem[Cencek \latin{et~al.}(2012)Cencek, Przybytek, Komasa, Mehl, Jeziorski,
  and Szalewicz]{cencek12}
Cencek,~W.; Przybytek,~M.; Komasa,~J.; Mehl,~J.~B.; Jeziorski,~B.;
  Szalewicz,~K. Effects of adiabatic, relativistic, and quantum electrodynamics
  interactions on the pair potential and thermophysical properties of helium.
  \emph{J. Chem. Phys.} \textbf{2012}, \emph{136}, 224303\relax
\mciteBstWouldAddEndPuncttrue
\mciteSetBstMidEndSepPunct{\mcitedefaultmidpunct}
{\mcitedefaultendpunct}{\mcitedefaultseppunct}\relax
\EndOfBibitem
\bibitem[Lehtola(2019)]{lehtola19}
Lehtola,~S. Fully numerical Hartree-Fock and density functional calculations.
  I. Atoms. \emph{Int. J. Quantum Chem.} \textbf{2019}, \emph{119},
  e25945\relax
\mciteBstWouldAddEndPuncttrue
\mciteSetBstMidEndSepPunct{\mcitedefaultmidpunct}
{\mcitedefaultendpunct}{\mcitedefaultseppunct}\relax
\EndOfBibitem
\bibitem[Nakashima and Nakatsuji(2008)Nakashima, and Nakatsuji]{nakashima08}
Nakashima,~H.; Nakatsuji,~H. Solving the electron-nuclear Schr{\"o}dinger
  equation of helium atom and its isoelectronic ions with the free
  iterative-complement-interaction method. \emph{J. Chem. Phys.} \textbf{2008},
  \emph{128}, 154107\relax
\mciteBstWouldAddEndPuncttrue
\mciteSetBstMidEndSepPunct{\mcitedefaultmidpunct}
{\mcitedefaultendpunct}{\mcitedefaultseppunct}\relax
\EndOfBibitem
\bibitem[Pachucki and Komasa(2004)Pachucki, and Komasa]{pachucki04}
Pachucki,~K.; Komasa,~J. Relativistic and QED corrections for the beryllium
  atom. \emph{Phys. Rev. Lett.} \textbf{2004}, \emph{92}, 213001\relax
\mciteBstWouldAddEndPuncttrue
\mciteSetBstMidEndSepPunct{\mcitedefaultmidpunct}
{\mcitedefaultendpunct}{\mcitedefaultseppunct}\relax
\EndOfBibitem
\bibitem[Prascher \latin{et~al.}(2011)Prascher, Woon, Peterson, Dunning, and
  Wilson]{prascher11}
Prascher,~B.~P.; Woon,~D.~E.; Peterson,~K.~A.; Dunning,~T.~H.; Wilson,~A.~K.
  Gaussian basis sets for use in correlated molecular calculations. VII.
  Valence, core-valence, and scalar relativistic basis sets for Li, Be, Na, and
  Mg. \emph{Theor. Chem. Acc.} \textbf{2011}, \emph{128}, 69--82\relax
\mciteBstWouldAddEndPuncttrue
\mciteSetBstMidEndSepPunct{\mcitedefaultmidpunct}
{\mcitedefaultendpunct}{\mcitedefaultseppunct}\relax
\EndOfBibitem
\bibitem[Przybytek and Lesiuk(2018)Przybytek, and Lesiuk]{przybytek18}
Przybytek,~M.; Lesiuk,~M. Correlation energies for many-electron atoms with
  explicitly correlated Slater functions. \emph{Phys. Rev. A} \textbf{2018},
  \emph{98}, 062507\relax
\mciteBstWouldAddEndPuncttrue
\mciteSetBstMidEndSepPunct{\mcitedefaultmidpunct}
{\mcitedefaultendpunct}{\mcitedefaultseppunct}\relax
\EndOfBibitem
\bibitem[Pavanello \latin{et~al.}(2009)Pavanello, Tung, Leonarski, and
  Adamowicz]{pavanello09}
Pavanello,~M.; Tung,~W.-C.; Leonarski,~F.; Adamowicz,~L. New more accurate
  calculations of the ground state potential energy surface of H$_3^+$.
  \emph{J. Chem. Phys.} \textbf{2009}, \emph{130}, 074105\relax
\mciteBstWouldAddEndPuncttrue
\mciteSetBstMidEndSepPunct{\mcitedefaultmidpunct}
{\mcitedefaultendpunct}{\mcitedefaultseppunct}\relax
\EndOfBibitem
\bibitem[Jensen(2005)]{jensen05}
Jensen,~F. Estimating the Hartree—Fock limit from finite basis set
  calculations. \emph{Theor. Chem. Acc.} \textbf{2005}, \emph{113},
  267--273\relax
\mciteBstWouldAddEndPuncttrue
\mciteSetBstMidEndSepPunct{\mcitedefaultmidpunct}
{\mcitedefaultendpunct}{\mcitedefaultseppunct}\relax
\EndOfBibitem
\bibitem[Bukowski \latin{et~al.}(1999)Bukowski, Jeziorski, and
  Szalewicz]{bukowski99}
Bukowski,~R.; Jeziorski,~B.; Szalewicz,~K. Gaussian geminals in explicitly
  correlated coupled cluster theory including single and double excitations.
  \emph{J. Chem. Phys.} \textbf{1999}, \emph{110}, 4165--4183\relax
\mciteBstWouldAddEndPuncttrue
\mciteSetBstMidEndSepPunct{\mcitedefaultmidpunct}
{\mcitedefaultendpunct}{\mcitedefaultseppunct}\relax
\EndOfBibitem
\bibitem[Barnes \latin{et~al.}(2008)Barnes, Petersson, Feller, and
  Peterson]{barnes08}
Barnes,~E.~C.; Petersson,~G.~A.; Feller,~D.; Peterson,~K.~A. The CCSD(T)
  complete basis set limit for Ne revisited. \emph{J. Chem. Phys.}
  \textbf{2008}, \emph{129}, 194115\relax
\mciteBstWouldAddEndPuncttrue
\mciteSetBstMidEndSepPunct{\mcitedefaultmidpunct}
{\mcitedefaultendpunct}{\mcitedefaultseppunct}\relax
\EndOfBibitem
\bibitem[Barnes and Petersson(2010)Barnes, and Petersson]{barnes10}
Barnes,~E.~C.; Petersson,~G.~A. MP2/CBS atomic and molecular benchmarks for H
  through Ar. \emph{J. Chem. Phys.} \textbf{2010}, \emph{132}, 114111\relax
\mciteBstWouldAddEndPuncttrue
\mciteSetBstMidEndSepPunct{\mcitedefaultmidpunct}
{\mcitedefaultendpunct}{\mcitedefaultseppunct}\relax
\EndOfBibitem
\bibitem[Flores(1993)]{flores93}
Flores,~J.~R. High precision atomic computations from finite element
  techniques: Second-order correlation energies of rare gas atoms. \emph{J.
  Chem. Phys.} \textbf{1993}, \emph{98}, 5642--5647\relax
\mciteBstWouldAddEndPuncttrue
\mciteSetBstMidEndSepPunct{\mcitedefaultmidpunct}
{\mcitedefaultendpunct}{\mcitedefaultseppunct}\relax
\EndOfBibitem
\bibitem[Flores(2008)]{flores08}
Flores,~J.~R. New benchmarks for the second-order correlation energies of Ne
  and Ar through the finite element MP2 method. \emph{Int. J. Quantum Chem.}
  \textbf{2008}, \emph{108}, 2172--2177\relax
\mciteBstWouldAddEndPuncttrue
\mciteSetBstMidEndSepPunct{\mcitedefaultmidpunct}
{\mcitedefaultendpunct}{\mcitedefaultseppunct}\relax
\EndOfBibitem
\bibitem[Thorpe \latin{et~al.}(2021)Thorpe, Kilburn, Feller, Changala, Bross,
  Ruscic, and Stanton]{thorpe21}
Thorpe,~J.~H.; Kilburn,~J.~L.; Feller,~D.; Changala,~P.~B.; Bross,~D.~H.;
  Ruscic,~B.; Stanton,~J.~F. Elaborated thermochemical treatment of HF, CO,
  N$_2$, and H$_2$O: Insight into HEAT and its extensions. \emph{J. Chem.
  Phys.} \textbf{2021}, \emph{155}, 184109\relax
\mciteBstWouldAddEndPuncttrue
\mciteSetBstMidEndSepPunct{\mcitedefaultmidpunct}
{\mcitedefaultendpunct}{\mcitedefaultseppunct}\relax
\EndOfBibitem
\bibitem[Przybytek \latin{et~al.}(2017)Przybytek, Cencek, Jeziorski, and
  Szalewicz]{przybytek17}
Przybytek,~M.; Cencek,~W.; Jeziorski,~B.; Szalewicz,~K. Pair Potential with
  Submillikelvin Uncertainties and Nonadiabatic Treatment of the Halo State of
  the Helium Dimer. \emph{Phys. Rev. Lett.} \textbf{2017}, \emph{119},
  123401\relax
\mciteBstWouldAddEndPuncttrue
\mciteSetBstMidEndSepPunct{\mcitedefaultmidpunct}
{\mcitedefaultendpunct}{\mcitedefaultseppunct}\relax
\EndOfBibitem
\bibitem[Karton and Martin(2021)Karton, and Martin]{karton21}
Karton,~A.; Martin,~J.~M. Prototypical $\pi$--$\pi$ dimers re-examined by means
  of high-level CCSDT(Q) composite ab initio methods. \emph{J. Chem. Phys.}
  \textbf{2021}, \emph{154}, 124117\relax
\mciteBstWouldAddEndPuncttrue
\mciteSetBstMidEndSepPunct{\mcitedefaultmidpunct}
{\mcitedefaultendpunct}{\mcitedefaultseppunct}\relax
\EndOfBibitem
\bibitem[Patkowski and Szalewicz(2010)Patkowski, and Szalewicz]{patkowski10}
Patkowski,~K.; Szalewicz,~K. Argon pair potential at basis set and excitation
  limits. \emph{J. Chem. Phys.} \textbf{2010}, \emph{133}, 094304\relax
\mciteBstWouldAddEndPuncttrue
\mciteSetBstMidEndSepPunct{\mcitedefaultmidpunct}
{\mcitedefaultendpunct}{\mcitedefaultseppunct}\relax
\EndOfBibitem
\bibitem[Pachucki and Sapirstein(2000)Pachucki, and Sapirstein]{pachucki00}
Pachucki,~K.; Sapirstein,~J. Relativistic and QED corrections to the
  polarizability of helium. \emph{Phys. Rev. A} \textbf{2000}, \emph{63},
  012504\relax
\mciteBstWouldAddEndPuncttrue
\mciteSetBstMidEndSepPunct{\mcitedefaultmidpunct}
{\mcitedefaultendpunct}{\mcitedefaultseppunct}\relax
\EndOfBibitem
\bibitem[Rychlewski(1980)]{rychlewski80}
Rychlewski,~J. An accurate calculation of the polarizability of the hydrogen
  molecule and its dependence on rotation, vibration and isotopic substitution.
  \emph{Mol. Phys.} \textbf{1980}, \emph{41}, 833--842\relax
\mciteBstWouldAddEndPuncttrue
\mciteSetBstMidEndSepPunct{\mcitedefaultmidpunct}
{\mcitedefaultendpunct}{\mcitedefaultseppunct}\relax
\EndOfBibitem
\bibitem[Martin(1996)]{martin96}
Martin,~J.~M. Ab initio total atomization energies of small molecules—towards
  the basis set limit. \emph{Chem. Phys. Lett.} \textbf{1996}, \emph{259},
  669--678\relax
\mciteBstWouldAddEndPuncttrue
\mciteSetBstMidEndSepPunct{\mcitedefaultmidpunct}
{\mcitedefaultendpunct}{\mcitedefaultseppunct}\relax
\EndOfBibitem
\bibitem[Varandas(2021)]{varandas21}
Varandas,~A. J.~C. Canonical and explicitly-correlated coupled cluster
  correlation energies of sub-kJ/mol accuracy via cost-effective
  hybrid-post-CBS extrapolation. \emph{Phys. Chem. Chem. Phys.} \textbf{2021},
  \emph{23}, 9571--9584\relax
\mciteBstWouldAddEndPuncttrue
\mciteSetBstMidEndSepPunct{\mcitedefaultmidpunct}
{\mcitedefaultendpunct}{\mcitedefaultseppunct}\relax
\EndOfBibitem
\bibitem[Lang \latin{et~al.}({})Lang, Przybytek, and Lesiuk]{github}
Lang,~J.; Przybytek,~M.; Lesiuk,~M. Extrapolation through random walk. {};
  \url{https://github.com/lesiukmichal/extrapolation-random-walk.git}\relax
\mciteBstWouldAddEndPuncttrue
\mciteSetBstMidEndSepPunct{\mcitedefaultmidpunct}
{\mcitedefaultendpunct}{\mcitedefaultseppunct}\relax
\EndOfBibitem
\end{mcitethebibliography}

\end{document}